\documentclass[twocolumn,6pt]{article}
\usepackage{balance}
\usepackage{etoolbox}
\usepackage{flushend}
\usepackage[section]{placeins}
\usepackage[bf, footnotesize]{caption}
\usepackage{graphics}
\usepackage{CJKutf8}

\usepackage{amssymb}
\usepackage{mathtools, cuted}
\usepackage{widetext}

\usepackage{lineno}

\usepackage{multicol}
\usepackage{booktabs}
\usepackage{amsmath}
\usepackage{widetext}
\usepackage{flushend}
\usepackage[explicit]{titlesec}

\titleformat{\section}
  {\bf}{\thesection}{1em}{\MakeUppercase{#1}}
  \titleformat{\subsection}
  {\it}{\thesection}{1em}{{#1}}
  
\begin{document}
\footnotesize
\begin{strip}

\title{Quantitative analysis of jugular venous pulse obtained by using a general-purpose ultrasound scanner}
\date{}


\author{Francesco Sisini\\
\begin{small}
Department of Physics and Earth Sciences, University of Ferrara, Via Saragat 1, 44122 Ferrara, Italy
\end{small}\\\\
\begin{tiny}
Published on 14 April 2016.
\end{tiny}
}


\begin{CJK}{UTF8}{min}\Large 拳必殺  notes series\end{CJK}

\begin{center}
\line(1,0){450}
\end{center}
\maketitle

\pagebreak


Ikken hissatsu (\begin{CJK}{UTF8}{min}拳必殺 )\end{CJK} means something like \textit{to annihilate at one blow}.
This document is part of a series of   notes each one targeting a single goal. Each note has  to \textit{annihilate  at one blow}! 

\section*{Conditions of use}

 This is a self-published methodological note distributed under the Creative Commons Attribution License (http://creativecommons.org/licenses/by/4.0/), which permits unrestricted use, distribution, and reproduction in any medium, provided the original work is properly cited. The note  contains an original  reasoning of mine and the goal to share thoughts and methodologies, not results. Therefore before using the contents of these notes, everyone is invited to verify the accuracy of the assumptions and conclusions.
\end{strip}

\section*{Introduction}
Jugular venous pulse (JVP) examination  dates back to more than a century ago \cite{Mackenzie,Mackay,Applefeld,Kalmanson}, but only recently it has  been shown that the trace of the JVP can be obtained by means of an ultrasound scanner (US) equipped with a commercial linear probe.\cite{sisini,sahani} 
The method proposed needs a video clip of a few seconds obtained from the transverse scan of the neck, where the internal jugular vein (IJV) is clearly visible. \cite{sisini} The cross sectional area of the IJV (CSA) is measured on each picture (sonogram) of the video clip. The sequence of measurements (data sets) is the CSA trace. Moreover, in the same studies, it was shown that acquiring the video clip of the IJV simultaneously to the ECG trace, allows to identify relationship between the two ECG and JVP traces (see Fig. \ref{fig:schemabase}).
The non-invasive techniques used up to now, are based on the use of a microphone or a motion sensor and produce a  trace which qualitatively represents the pressure in the IJV.\cite{Pyhel,Applefeld,Mackenzie,Mackay,Kalmanson}
This new technique is different from those used previously in that the JVP is the instantaneous value of the CSA of the IJV. The US technique for the  JVP represents an improvement from qualitative to quantitative JVP. The importance of having a quantitative JVP trace had already emerged in the past when a calibrated JVP has been used.\cite{Pyhel} This new technique goes beyond the limit of calibration because it makes it possible to extrapolate the numeric parameters from the JVP that can be used for the clinical evaluation inter and intra-patient.
This is a methodology-note that describes the innovative methods developed by the author to obtain quantitative parameters obtained by the JVP with ultrasound technique.  The results reported here are only illustrative and are not produced by a dedicated experiment.

\section*{Methodology presentation} 
\subsection*{ ECG analysis and P, R e T events detection}
\begin{figure}
\includegraphics[scale=0.3]{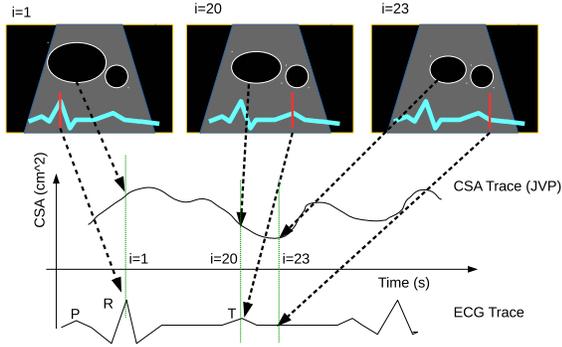} 
\caption{ The figure represents an illustrative sequence of three sonograms from a video-clip obtained by a transverse ultrasound scan of the IJV. In each picture it is visible the IJV (ellipses), the common carotid artery (circular) and the ECG trace (blue line). The active phase of the ECG is indicated by the red cursor. Below, it shows the IJV  CSA trace obtained by measuring the CSA in cm $ ^ 2 $ on each sonogram of the  video-clip }
\label{fig:schemabase}
\end{figure}

Over the past 30 years,    several algorithms for the automatic detection of structures P, QRS and T in ECG, have been developed\cite{Kohler}. However, these can also be analysed manually.
In this work, the video clip of the IJV was analyzed frame by frame to find the sonograms in which the cursor of the ECG was  at on of  the events P, R or T (see Fig. \ref{fig:schemabase}). The time instants relating to each event were defined $ t_ {P_ {i}} $, $ t_ {R_ {i}} $ and $ t_ {T_ {i}} $, where $ i $ indicates the $ i $-th cardiac cycle between those traces.
\subsection*{JVP quantitative evaluation}
\begin{figure}

\includegraphics[scale=0.3]{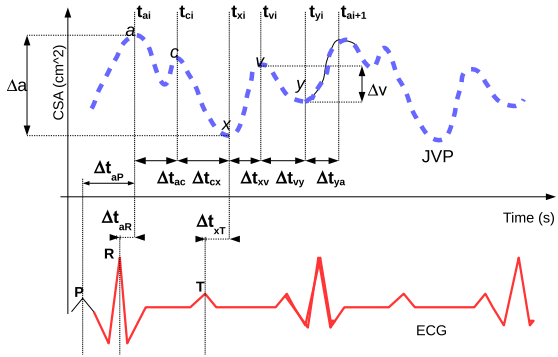} 
\caption{The figure shows an illustrative  JVP trace during a cardiac cycle together with the ECG trace. The five waves "a", "c", "x", "v" and "y" of the JVP and the  events P, R and T of the ECG are indicated.} 
\label{fig:schema}
\end{figure}

\begin{figure}

\includegraphics[scale=0.2]{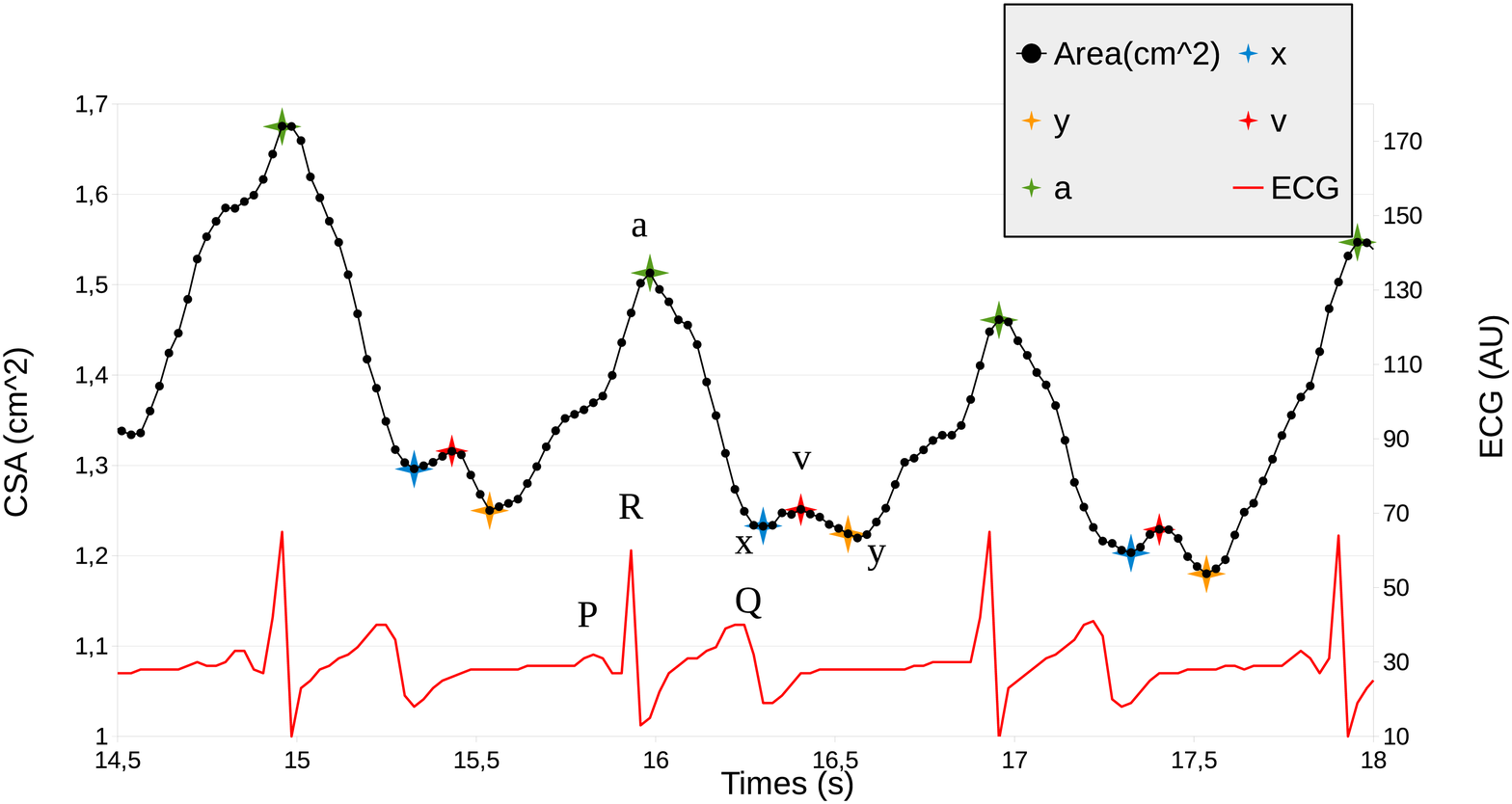} 
\caption{Examples of CSA and ECG trace are shown. The waves a,  \textit{x}, \textit{v} e \textit{y} are correctly detected by  an algorithm. The P, R e T events of the ECG trace are indicated.
}
\label{fig:jvp}
\end{figure}

\begin{figure}
\includegraphics[scale=0.2]{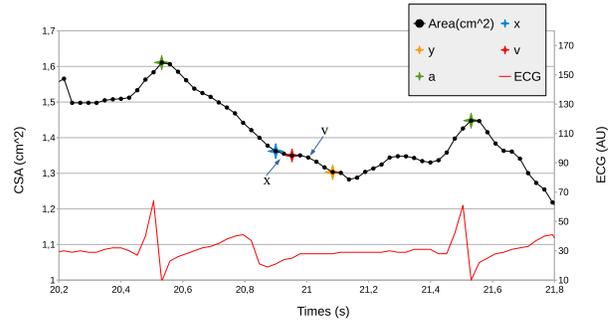} 
\caption{
Examples of CSA and ECG trace are shown. The   \textit{x} and  \textit{v} e \textit{y} are \textbf{not}  correctly detected. Correct waves are indicated. The P, R e T events of the ECG are indicated.}
\label{fig:jvp_we}
\end{figure}

  The waves $ a$, $ c $, $ x $, $ v $ and $ y $ and the intervals $ \Delta a$ and $ \Delta v $ are the parameters that characterize the JVP (see Fig. \ref{fig:schema}). In this paper, these parameters are defined in terms of the values assumed by the CSA during the cardiac cycle. The values of the CSA of the IJV in correspondence of the waves $ a $, $ c $, $ x $, $ v $ and $ y $, during the $i$-th cycle are indicated by the parameters  $a_i$, $c_i$, $x_i$, $v_i$ and $y_i$.
These parameters are detected on the JVP trace by means of an algorithm based on the following definitions:

\begin{widetext}
\begin{equation}
\label{eqn:int}
\begin{array}{r}
a_1=max(CSA(t)): 0<t<T_c\\
a_i=max(CSA(t)): t_{y_{i-1}}+\Delta  t_{ya}-\gamma T_c<t_{a_{i}}<t_{y_{i-1}}+\Delta  t_{ya}+\gamma T_c\\
c_i=max(CSA(t)): t_{a_{i}}+\Delta t_{ac}-\gamma T_c<t<t_{a_{i}}+\Delta t_{ac}+\gamma T_c\\
x_i=min(CSA(t)): t_{a_{i}}+\Delta t_{ax}-\gamma T_c<t<t_{a_{i}}+\Delta t_{ax}+\gamma T_c\\
v_i=max(CSA(t)): t_{x_{i}}+\Delta t_{xv}-\gamma T_c<t<t_{x_{i}}+\Delta t_{cx}+\gamma T_c\\
y_i=min(CSA(t)): t_{v_{i}}+\Delta t_{vy}-\gamma T_c<t<t_{v_{i}}+\Delta t_{vy}+\gamma T_c
\end{array}
\end{equation}
\end{widetext}
The parameter $\gamma$ ranges  between 0 and 1 (typically 0.05) and serves to take account of  heart rate variation (HRV) during acquisition.
The $ t_{a_{i-1}} $ is the instant corresponding to the wave \textit{a} during the $ (i-1)$-th cycle, the parameters $ t_ {x_ {i}} $ and $ t_ {v_ {i}} $ are defined similarly.
The parameters $ \Delta t_ {ac} $, $ \Delta t_ {ax} $, $ \Delta t_ {xv} $ and $ \Delta t_ {vy} $ represent the time interval between the waves \textit{a} and \textit{c} , \textit{c} and \textit{x}, \textit{x} and \textit{v},

\textit{v} and \textit{y}, respectively (see Fig. \ref{fig:schema}) and are measured by an operator over a   selected cardiac cycle. The detection algorithm takes into account  their possible modification during susequent cycles, by means of the parameter $\gamma$ (see Eq. (\ref{eqn:int})). Once
identified the waves $a_i$, $c_i$, $x_i$, $v_i$ e $y_i$ on each cardiac cycle,  the intervals $ \Delta t_ {ax_i} $ and $ \Delta t_ {vy_i} $ are calculated for each cardiac cycle. The parameters defined in Eq. \ref{eqn:int} are used to define the two quantities

\begin{equation}
\label{eqn:deltaa}
\begin{array}{r}
\Delta a_i=a_i-x_i\\
\Delta v_i=v_i-y_i\\
\end{array}
\end{equation}
The average and the standard deviation for e each parameter $a_i$, $v_i$, $\Delta a_i$, $\Delta v_i$,  $\Delta t_{ax_i}$ and $\Delta t_{vy_i}$   are calculated over several cardiac cycle.\\\\
An illustrative JVP trace is presented in Fig.\ref{fig:jvp} where the waves  \textit{a}, \textit{c}, \textit{x}, \textit{v} and \textit{y} are visible.
The parameters $a_i$, $x_i$, $v_i$ and $y_i$ for each cardiac cycle $i $ were identified by the algorithm. However, about 10 \% of the points are normally not properly detected by the algorithm, while it was possible to identify them manually (see Fig. \ref{fig:jvp_we}).
The parameters $ \Delta a_i $, $ \Delta v_i $, $ \Delta t_ {ax_i} $ and $ \Delta t_ {vy_i} $ were calculated as in Eq. (\ref{eqn:deltaa}), The mean values and standard deviations of the parameters are presented in Table \ref{tab:parameters}.
\begin{table*}[]
\tiny

\caption{Example of mean value and standard deviation of main JVP parameters  are presented.}
\resizebox{\textwidth}{!}{%
\label{tab:parameters}
\begin{tabular}{llllllll}
 a     &x                & v                     & $\Delta$a                    & $\Delta$v                                        & $\Delta$t$_{ax}$     & $\Delta$t$_{vy}$     & HR  \\
         (cm\textasciicircum 2)  & (cm\textasciicircum 2) & (cm\textasciicircum 2) & (cm\textasciicircum 2) & (cm\textasciicircum 2) &  (s)       & (s)       & Bpm \\
 1.57 $\pm$ 0.18   &1.29$\pm$0.16           & 1.33$\pm$0.18               & 0.29$\pm$0.04              & 0.04$\pm$0.02                          & 0.33$\pm$0.06 & 0.11$\pm$0.05 & 61  
\end{tabular}
}
\end{table*}
\begin{table*}[]
\centering
\caption{Example of mean value and standard deviation of of time intervals  $\Delta$t$_{aR}$ and $\Delta$t$_{xT}$ averaged over several cardiac cycles.}
\label{tab:ecg}
\begin{tabular}{lll}
         $\Delta$t$_{aR}$      & $\Delta$t$_{xT}$ &  $\Delta$t$_{aP}$      \\
 (s)       & (s) & (s)      \\
 0,03$\pm$ 0,02 & 0,12$\pm$ 0,06 &  0.14 $\pm$ 0.05   \\

\end{tabular}
\end{table*}
\subsection*{ JVP and ECG time relationship }
The  time interval between the wave \textit{a} and the events P and R and the wave \textit{x} and the event T are calculated for each cardiac cycle. These intervals are defined as:
\begin{equation}
\label{eqn:deltajvpecg}
\begin{array}{r}
\Delta t_{aR_i}=t_{a_i}-t_{R_i}\\
\Delta t_{aP_i}=t_{a_i}-t_{P_i}\\
\Delta t_{xT_i}=t_{x_i}-t_{T_i}\\
\end{array}
\end{equation}
The average and the standard deviation of these three parameters is calculated. The mean values and standard deviations of the parameters $ \Delta t_ {aR} $, $ \Delta t_ {xT} $ and $ \Delta t_ {aP} $ are presented in Table \ref{tab:ecg}.

\section*{Discussion}
This note describes a methodology to extrapolate some interesting quantitative parameters from the IJV CSA trace, which is an analogue of the JVP.
There are two important aspects  that must be considered: i) the contouring of the IJV on each sonogram of the video-clip must be as accurate as possible. This means that the algorithm presented in \cite{sisini} must be improved to achieve an accuracy close to 100 \%, while the current one \ 'and about 95 \%. ii) The wave detection algorithm must be improved. In fact it is not always sufficient to identify the waves, while they can be identified by the analysis of tracks performed by a human expert.
This paper has not discussed the clinical use of these parameters.
\section*{Acknowledgement}
The author wants to thank Giacomo Gadda, Valentina Tavoni and Valentina Sisini to have kindly review the manuscript.




\pagebreak

\end{document}